\documentstyle[epsfig,aps,prd]{revtex}
\begin{document}
\twocolumn[\hsize\textwidth\columnwidth\hsize\csname
 @twocolumnfalse\endcsname
\title{Cosmology with tachyon field as dark energy}
\author{J.~S.~Bagla$^a$\thanks{jasjeet@mri.ernet.in},
  H.~K.~Jassal$^b$\thanks{hkj@iucaa.ernet.in},
  T.~Padmanabhan$^b$\thanks
{nabhan@iucaa.ernet.in}}  
\address{$^a$ Harish-Chandra Research Institute, Chhatnag Road,  
Jhunsi, Allahabad-211 019, India. \\
$^b$ Inter-University Centre for Astronomy and Astrophysics, Post Bag
4, Ganeshkhind, Pune-411 007, India.}  
\maketitle
\begin{abstract} 
We present a detailed study of cosmological effects of homogeneous
tachyon matter coexisting with non-relativistic matter and radiation,
concentrating on the 
  inverse square potential and the exponential potential for the
  tachyonic scalar field. 
A distinguishing feature of these models (compared to other cosmological
models) is that the matter density parameter and the density parameter
for tachyons remain comparable even in the matter dominated phase.  
For the exponential potential, the solutions have an accelerating phase,
{\it followed by} a phase with $a(t)\propto t^{2/3}$ as $t\to
\infty$. 
This eliminates the future event horizon present in $\Lambda CDM$ models
and is an attractive feature from the string theory perspective.
A comparison with supernova Ia data  shows that for both the potentials
there exists a range of models in which the universe
undergoes an accelerated expansion at low redshifts and are also
consistent with requirements of structure formation. 
They do require fine tuning of parameters but not any more than in the
case of $\Lambda CDM$ or quintessence models. 
\end{abstract}
]
\section{Motivation}
Observations suggest that our universe has entered a phase of 
accelerated expansion  in the recent past. 
Friedmann equations can be consistent with such an accelerated
expansion only if the universe is populated by a medium with
negative pressure. 
One of the possible sources which could provide such a negative
pressure will be a scalar field with either of the following two types
of Lagrangians:
\begin{eqnarray}
L_{\rm quin} &=& \frac{1}{2} \partial_a \phi \partial^a \phi - V(\phi);
\\ \nonumber 
\quad L_{\rm tach}
&=& -V(\phi) [1-\partial_a\phi\partial^a\phi]^{1/2}
\end{eqnarray}
  Both these Lagrangians involve one arbitrary function $V(\phi)$. The first one
  $L_{\rm quin}$,  which is a natural generalization of the Lagrangian for
  a nonrelativistic particle, $L=(1/2)\dot q^2 -V(q)$, is usually called quintessence (for
  a sample of models, see Ref. \cite{phiindustry}).
    When it acts as a source in Friedmann universe, it is characterized
    by a time dependent   $w(t)\equiv (P/\rho)$  with
  %
  %
\begin{eqnarray}
  \rho_q(t) &=& \frac{1}{2} \dot\phi^2 + V; \quad P_q(t) = \frac{1}{2};
  \dot\phi^2 - V; \\ \nonumber 
w_q &=& \frac{1-(2V/\dot\phi^2)}{1+ (2V/\dot\phi^2)}.
  \label{quintdetail}
  \end{eqnarray}

Just as $L_{\rm quin}$ generalizes the Lagrangian for the nonrelativistic 
particle, $L_{\rm tach}$ generalizes the Lagrangian for the relativistic particle \cite{tptirth}.
 A relativistic particle with a (one dimensional) position
$q(t)$ and mass $m$ is described by the Lagrangian $L = -m \sqrt{1-\dot q^2}$.
It has the energy $E = m/  \sqrt{1-\dot q^2}$ and momentum $p = m \dot
q/\sqrt{1-\dot q^2} $ which are related by $E^2 = p^2 + m^2$.  As is well
known, this allows the possibility of having massless particles with finite
energy for which $E^2=p^2$. This is achieved by taking the limit of $m \to 0$
and $\dot q \to 1$, while keeping the ratio in $E = m/  \sqrt{1-\dot q^2}$
finite.  The momentum acquires a life of its own,  unconnected with the
velocity  $\dot q$, and the energy is expressed in terms of the  momentum
(rather than in terms of $\dot q$)  in the Hamiltonian formulation. We can now
construct a field theory by upgrading $q(t)$ to a field $\phi$. Relativistic
invariance now  requires $\phi $ to depend on both space and time [$\phi =
\phi(t, {\bf x})$] and $\dot q^2$ to be replaced by $\partial_i \phi \partial^i
\phi$. It is also possible now to treat the mass parameter $m$ as a function of
$\phi$, say, $V(\phi)$ thereby obtaining a field theoretic Lagrangian $L =-
V(\phi) \sqrt{1 - \partial^i \phi \partial_i \phi}$. The Hamiltonian  structure of this
theory is algebraically very similar to the special  relativistic example  we
started with. In particular, the theory allows solutions in which $V\to 0$,
$\partial_i\phi \partial^i\phi \to 1$ simultaneously, keeping the
energy (density) finite.  Such 
solutions will have finite momentum density (analogous to a massless particle
with finite  momentum $p$) and energy density. Since the solutions can now
depend on both space and time (unlike the special relativistic example in which
$q$ depended only on time), the momentum density can be an arbitrary function
of the spatial coordinate. 
This form of scalar field arises  in string theories \cite{asen} and --- for technical reasons ---
   is called a tachyonic scalar field.
   (The structure of this Lagrangian is similar to those analyzed in a wide class of models
   called {\it K-essence}; see for example, Ref. \cite{armen}.)
This provides a rich gamut of possibilities in the
context of cosmology
 \cite{tptirth,armen,cosmos,earl,tptachyon}.

   The stress tensor for the tachyonic scalar  field can be written in a
perfect fluid form
\begin{equation}
T^i_k = (\rho + p) u^i u_k - p \delta^i_k
\end{equation}
with
\begin{eqnarray}
u_k &=& \frac{\partial_k \phi}{\sqrt{\partial^i \phi \partial_i \phi}};\quad 
\rho = \frac{V(\phi)}{\sqrt{1 - \partial^i \phi \partial_i
    \phi}};\quad  \\ \nonumber
p &=& -V(\phi) \sqrt{1 - \partial^i \phi \partial_i \phi}
\end{eqnarray}
The remarkable feature of this stress tensor is that it could be considered
as {\it the sum of a pressure less dust component and a cosmological constant} \cite{tptirth}.
To show this explicitly,
we  break up the density $\rho$ and the pressure $p$
and write them in a more suggestive form as
\begin{equation}
\rho = \rho_V  + \rho_{\rm DM}; ~~
p = p_V  + p_{\rm DM},
\end{equation}
\\
where
\begin{eqnarray}
\rho_{\rm DM} &=& \frac{V(\phi) \partial^i \phi \partial_i \phi}
{\sqrt{1 - \partial^i \phi \partial_i \phi}};\ \  p_{\rm DM} = 0; \\ \nonumber
\rho_V  &=& V(\phi) \sqrt{1 - \partial^i \phi \partial_i \phi};\ \
p_V  = -\rho_V
\end{eqnarray}
This means that the stress tensor can be thought of as made up of two components
-- one behaving like a pressure-less fluid, while the other having a negative
pressure. In the cosmological context, when $\dot\phi$ is small
(compared to $V$ in the case of quintessence or compared to unity in
the case of tachyonic field), both these sources have $w\to -1$ and
mimic a cosmological constant. When $\dot \phi \gg V$, the
quintessence has $w\approx 1$ leading to    $\rho_q\propto (1+z)^6$;
the tachyonic field, on the other hand, has $w\approx 0$ for
$\dot\phi\to 1$    and behaves like non-relativistic matter.
  
An additional motivation for studying models based on $L_{\rm quin}$ 
is the following:  
The standard explanation of the current cosmological observations
will require two components of dark matter: (a) The first one is a dust component
with the  equation of state  $p=0$ contributing $\Omega_m \approx 0.35$. This
component clusters gravitationally at small scales  ($l \lesssim 500$ Mpc, say)
and will be able to explain observations from galactic to super-cluster scales.
(b) The second one is a negative pressure component with equation of state like
$p=w\rho$  with $-1 < w < -0.5 $ contributing about $\Omega_V \approx 0.65$.
There is  some leeway in the $(p/\rho)$ of the second component but it is
certain that  $p$ is negative and $(p/\rho)$ is of order unity. 
The cosmological constant will provide $w=-1$ while
several other candidates based on scalar fields with potentials
\cite{phiindustry} will provide different values for $w$ in the acceptable
range.  By and large, component (b) is noticed only in the large scale
expansion and it does not cluster gravitationally to a significant extent.
Neither of the components (a) and (b) has laboratory evidence for their 
existence directly or indirectly. In this sense, cosmology requires invoking
the tooth fairy twice to explain the current observations. It was suggested
recently \cite{tptirth} that one may be able to explain the observations at all
scales using a single scalar field with a particular form of Lagrangian.

In this paper we explore the cosmological scenario in greater detail, concentrating
on the background cosmology.
   Our approach in this paper will be based on the above view point and we shall
   treat the form of the Lagrangian for $L_{\rm tach}$ as our starting
   point without worrying about its    origin. In particular, we shall
   {\it not}     make any attempt to connect up the form of
   $V(\phi)$ with string theoretic models but will explore different
   possibilities,    guided essentially by their cosmological viability. 
   

We construct cosmological models with homogeneous tachyon matter,
assuming that tachyon matter co-exists with normal non-relativistic
matter and radiation.      
Section \ref{sec:tachmatter} presents the equations for the evolution of scale factor
and the tachyon field. 
In Section \ref{sec:backgd_solns} we present solutions of these equations and discuss
variations introduced by the available parameters in the model.
Here we have analyzed two different models of the scalar field
potential, one is the exponential potential and the other is the the
inverse square potential which leads to power law cosmology
\cite{tptachyon}.   
In Section \ref{sec:supnova_data} we compare the model with observations and constrain
the parameters.
Section \ref{sec:tachstr} discusses structure formation in tachyon models.
The results are summarized in concluding Section \ref{sec:conclusions}.

\section{Tachyon-Matter cosmology}
\label{sec:tachmatter}

For a spatially flat universe, the Friedman equations are 
\begin{eqnarray}
\label{eq:frw}
\left(\frac{\dot a}{a}\right)^2 = \frac{8 \pi G}{3} \rho,~~~~
\frac{\ddot{a}}{a}= -\frac{4 \pi G}{3} (\rho + 3p) \\ \nonumber
\end{eqnarray}
where $\rho=\rho_{\mathrm NR} + \rho_{\mathrm R} + \rho_{\phi}$, with
respective terms denoting non-relativistic, relativistic and tachyon
matter densities. 
For the tachyon field $\phi$ we have
\begin{eqnarray}
\label{eq:tach}
\rho_{\phi}=\frac{V(\phi)}{\sqrt{1-\dot{\phi}^2}},~~~~
p_{\phi} = -V(\phi) \sqrt{1-\dot{\phi}^2} \\ \nonumber
\end{eqnarray}
\noindent The equation of state for tachyon matter is $p=w\rho$ with
$w=\dot{\phi}^2 - 1$.
The scalar field equation of motion is    
\begin{equation}
\label{eq:sc_motion}
\ddot{\phi} =  - (1-\dot{\phi}^2)  \left[ 3H \dot{\phi} +
\frac{1}{V(\phi)} \frac{dV}{d \phi}\right]
\end{equation}
The structure of this equation suggests that the change in $\dot\phi$
goes to zero as it approaches $\pm 1$.  
In this case, the equation of state for the tachyon field is dustlike.
Thus at any stage if the tachyon field behaves like dust, it will
continue to do so for a long time.  
This behavior persists for a duration that depends on the closeness
of $\dot\phi$ to $\pm 1$.  
Detailed behavior will depend on the form of the potential.

We shall discuss cosmological models with two different $V(\phi)$ in
this paper. 
The first one has the form 
$$
V(\phi) =\frac{n}{4 \pi G} \left(1 - \frac{2}{3n}\right)^{1/2} \phi^{-2}.
$$  
It was shown in  Ref. \cite{tptachyon} that the above potential
leads to the expansion $a(t)=t^n$ if $\phi$ is the only source.
(The  form $V(\phi) \propto 1/(\phi-\phi_0)^2$   can be
reduced to the form given above by a simple redefinition of the scalar
field.)  
In this case, the term in the square bracket in equation
(\ref{eq:sc_motion}) is $\left[3 H \dot\phi - 2/\phi \right]$.    
This term vanishes in the asymptotic limit when tachyons dominate and $\phi
\propto t$.  
This asymptotic solution is stable in the sense that all initial
conditions eventually lead to this state. 
If normal matter or radiation dominates, $\dot\phi$ stays close to the
transition point $2/(3H\phi)$, unless we start the field very close to
$\dot\phi^2 = 1$.   
Thus the presence of non-relativistic matter and radiation leads to a
change in the equation of state for the tachyon field.
The change in equation of state for the
tachyon field implies that it is {\it not} a tracker field (for details see
\cite{chiba1,chiba2}). 
To get a viable model, i.e., matter domination at high redshifts and
an accelerating phase at low redshifts, we need to start the tachyon
field such that $\dot\phi$ is very close to unity, and $\phi$ is very
large.   
We will describe the fine tuning required in greater detail when we
discuss numerical solutions of these equations.

The second form of potential which we will consider is the
exponential one with $V(\phi) \propto e^{-\phi/\phi_0}$. 
Then the  term in square bracket in the  equation (\ref{eq:sc_motion})
can be written as $\left[3H\dot\phi - 1/\phi_0 \right]$.     
In a universe dominated by radiation or normal matter, $H(t)$ is a
monotonically decreasing function of time while  $\phi_0$ is a constant.   
From this equation, it is clear that $\dot\phi$ will increase if
$\dot\phi < 1/(3H(t)\phi_0)$.  
As $H(t)$ keeps decreasing, $\dot\phi$ will increase
slowly and approach unity, asymptotically. 
However, in the meanwhile, tachyons may begin to dominate in terms
of energy density and this changes the behavior.  
In a tachyon dominated scenario, we get rapid expansion of the
universe and $H(t)$ varies much more slowly than in the matter or
radiation dominated era.  
Thus $\dot\phi$ changes at a slower rate but it still approaches unity
and hence we --- eventually --- get a dust like equation of state for the
tachyon field. 
Whether there is an accelerating phase or not depends on the initial
values of $\dot\phi$, $\phi_0$ and $\Omega_\phi$.
Present values of density parameter for non-relativistic matter and
tachyons fixes the epoch at which tachyons start to dominate the
energy density.
The parameter $\phi_0$ sets the time when $\dot\phi$ approaches unity
and the asymptotic dust like phase for tachyons is reached.  
Initial value of $\dot\phi$ fixes the duration of the accelerating
phase.  
If this value is very close to unity then it departs very little from
this value through the entire evolution.
On the other hand, if it starts far away from this value, then the
equation of state for the tachyon field can lead to a significantly
long accelerating phase. 
It is possible to fine tune the evolution  by choosing a sufficiently large
value for the constant $\phi_0$, so that $\dot\phi$ is much smaller than
unity even at the present epoch, and by requiring that the
tachyon field is starting to dominate the energy density of the
universe at the present epoch.  
In such a case, the accelerating phase is a transient between the
matter dominated era and the tachyon dominated era with $a(t) \propto
t^{2/3}$ in both of these regimes.  
This model has been the focus of many studies \cite{expo1}
though most of these have chosen to ignore the role of matter or
radiation. 

Some of the models discussed below have an accelerating phase for the universe
\emph{followed by} a dust-like expansion with $a(t) \propto t^{2/3}$
asymptotically. 
These models have the attractive feature that they do \emph{not} behave as a 
de Sitter-like universe in the asymptotic future and thus do not
possess a future horizon. 
String theoretic models have difficulty in incorporating cosmologies
which approach de Sitter-like phase asymptotically; from this  
point of view, these models can be theoretically \hfill  \linebreak attractive.


\section{Numerical Solutions to cosmological equations} 
\label{sec:backgd_solns}
This section presents a detailed study of the background tachyon
cosmology. 
We solve the cosmological equations of motion numerically for the two
potentials mentioned in the previous section. 

\subsection{Inverse square potential}
\label{inv_square}
We start with the discussion of the potential $V(\phi) \propto \phi^{-2}$. 
Energy density of the universe is a mix of the tachyon field,
radiation and non-relativistic matter.  
The asymptotic solutions for the universe with only the  tachyon field
as source  are well understood: we  get rapid expansion of the
universe with the field $\phi$ growing in proportion with time.  
We wish to see the effect of other species (matter and
radiation) on the evolution of the tachyon field, and the stability of
the asymptotic solution. 

Since only two of the three equations in (\ref{eq:frw}) and
(\ref{eq:sc_motion}) are independent, we choose to drop the
equation with second derivative of the scale factor.
We choose some instant of time as $t_{\rm in}$ and  rescale variables
as follows, for numerical convenience: 
\begin{equation}
x=t H_{in},~~
y=a(t)/a(t_{in}),~~
y'=dy/dx
\end{equation}
 and
$\psi=\phi/\phi_{in},\
\psi'=d\psi/dx$.
(We will use the present epoch and Planck epoch for $t_{\rm in}$, in
two different contexts described below.) 
Then
 the equations are
\begin{eqnarray}
y' &=& y \left[\frac{\Omega_{M_{in}}}{y^3}
  +\frac{\Omega_{R_{in}}}{y^4}+\frac{2 n}{3}  \left(1 - 
\frac{2}{3n}\right)^{1/2} \right. \\ \nonumber 
& &~~~~\times ~~~\left.\frac{1}{ \phi_{in}^2 H_{in}^2 \psi^2 \sqrt{1-
\phi_{in}^2 H_{in}^2 \psi'^2}} \right]^{1/2} \\ 
\psi'' &=& \left(1 - \phi_{in}^2 H_{in}^2 \psi'^2\right)\left(
 \frac{2}{\phi_{in}^2 H_{in}^2 \psi} -3 \frac{y'}{y} \psi'_{in} \right)
\end{eqnarray}
The subscript {\it in} refers to the initial value, and $\Omega_M$ and
$\Omega_R$ refer to the density parameters for matter and radiation
respectively.
It is assumed that the sum of all density parameters is unity.

The initial conditions for $y$ and $\psi$ follow from their definition:
$y_{in} = 1,\  
\psi_{in} = 1 $.
The initial condition for $\psi'$,  can be related to the
density parameter of tachyon matter at $t=t_{\rm in}$
\begin{eqnarray}
\Omega_{{\phi}_{in}} &=& \frac{2n}{3}\left(1 -
  \frac{2}{3n}\right)^{1/2}\frac{1}{\phi_{in}^2 H_{in}^2 
    \psi_{in}^2\sqrt{1-\phi_{in}^2 H_{in}^2 \psi'^2_{in}}}
\end{eqnarray}

\begin{figure}
\begin{center}
\epsfig{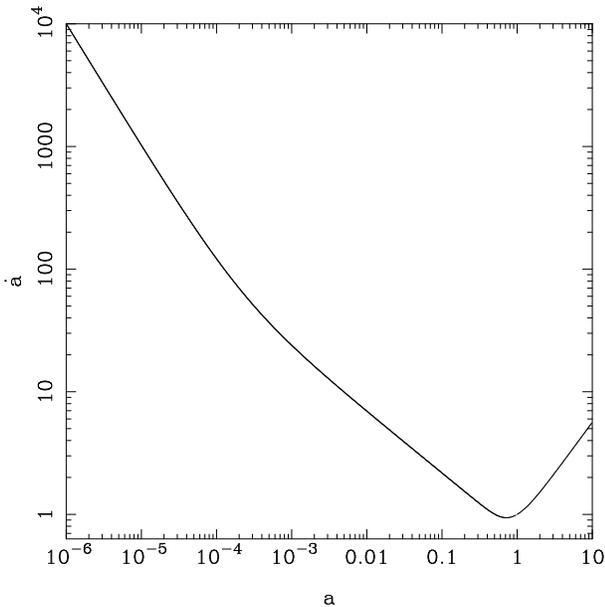}
\end{center}
\caption{Phase portrait for the scale factor.  This plot is for $\Omega_M
(present)=0.3$ and $n=6$. We choose the present value of $\phi_{in}
H_{in}=2.4$ where `in' refers to present epoch $(a=1; z=0)$. The transition from a decelerated 
expansion to an accelerated one is clear in this picture. We have used
this model for all the figures pertaining to the inverse square
potential.}
\label{fig:phase}
\end{figure}

There are two branches of solutions, one with positive $\psi'$ and
another with negative values of $\psi'$. 
For each set of $\Omega_{m_{in}}$ and $n$, there is a minimum value of 
$\phi_{in} H_{in}$ below which we cannot satisfy this equation.  
We can solve these equations if $\psi'^2 > 0$, or if
\begin{equation}
\phi_{in}^2 H_{in}^2 \geq \frac{2n}{3 \Omega_{{\phi}_{in}}} \left( 1 -
  \frac{2}{3n} \right)^{1/2}. 
\end{equation}

For a given value of $\phi_{in}^2 H_{in}^2$, we
get initial values of all the variables from the above equations and
using these we evolve these quantities.  
As shown in above equations, parameters $n$, $\Omega_{{\phi}_{in}}$,
$\phi_{in} H_{in}$ and $|\dot{\phi}_{in}|$ are inter-related.  
For a fixed $n$ and $\Omega_{{\phi}_{in}}$, larger values of
$\phi_{in}^2 H_{in}^2$ imply a larger value of initial $|\dot{\phi}|$,
and hence an equation of state that is closer to that of dust. 

We study the evolution of such a model in three steps.  
First we evolve the system from present day to future and show that the
asymptotic solution\cite{tptachyon} is stable, i.e., the entire
allowed range of present values leads to the same asymptotic solution.
In this context '{\it in}' refers to the present values of the
parameters. 
We choose the present value of $\Omega_{M_{in}}=0.3$, and all the
results shown here use $n=6$ (results do not change qualitatively for
other values of $n$), i.e., the asymptotic solution gives $a(t)
\propto t^6$.  

\begin{figure}
\begin{center}
\epsfig{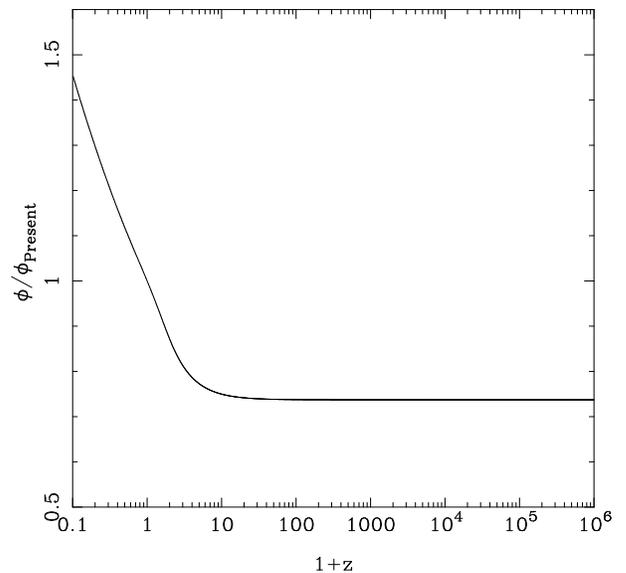}
\end{center}
\caption{The tachyon field $\phi/\phi_{\mathrm Present}$ as a function of
  redshift. The model is the same as in Fig. 1}
\label{fig:phi}
\end{figure}

\begin{figure}
\begin{center}
\epsfig{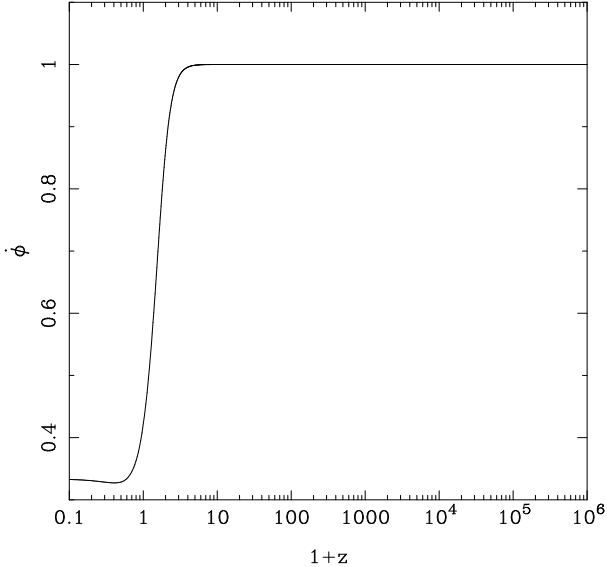}
\end{center}
\caption{This plot shows $\dot{\phi}$ as a function of redshift for the
  same model.}
\label{fig:phidot}
\end{figure}

\begin{figure}
\begin{center}
\epsfig{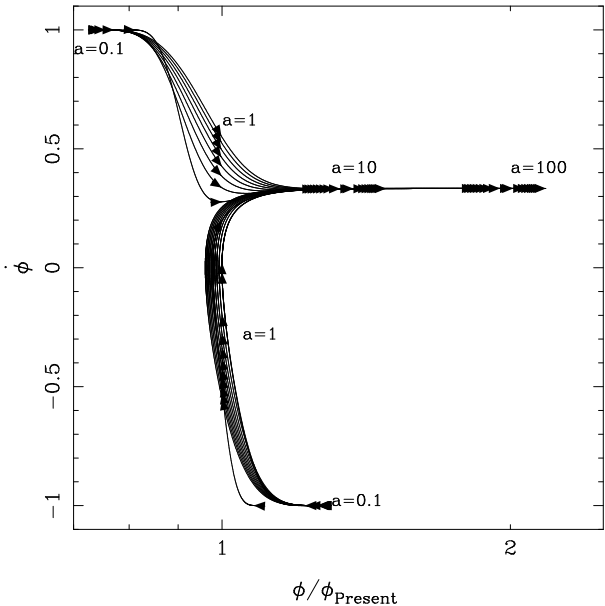}
\end{center}
\caption{
Phase plot for the tachyon field.  Here, we started from
initial conditions with a fixed value of $\phi$ and $n$.  $\dot\phi$
was varied and $\Omega_{\phi_{in}}$ was allowed to vary with it. Both
the positive branch as well as the negative branch are shown. The
scale factor $a(t)=1/(1+z)$ is marked along the arrow heads.} 
\label{fig:phiphase}
\end{figure}

We plot the phase portrait for the scale factor in
Fig.~\ref{fig:phase}.   
This figure clearly shows the late time acceleration in tachyon
dominated universe. (The figures exhibit past as well as future
evolution; the past evolution is discussed later.) 
The positive branch is plotted in this figure and the following ones but 
the discussion holds true for the negative branch as well.
Evolution of field $\phi$ and its time derivative $\dot\phi$ is shown
in Figs.~\ref{fig:phi} and ~\ref{fig:phidot}.  
These too demonstrate the approach to the asymptotic solution for the
two initial conditions. 
In asymptotic future, we expect $\dot\phi$ to approach a constant
value of $\sqrt{2/3n}$.   
Indeed, both the negative and the positive branches (which
respectively correspond to positive $\psi'$ and negative $\psi'$)
exhibit this behavior.  
The fact that other initial conditions lead to similar behavior is
shown in the phase diagram for field $\phi$ in
Fig.~\ref{fig:phiphase}.   
We see that for the entire range of initial conditions for the field
$\phi$, it quickly approaches  the asymptotic solution.  
It takes longer to reach the asymptotic solution for initial
conditions where $|\phi_{in} H_{in}|$ (the value at present) is much
larger than its minimum allowed value as in this case $|\dot\phi|$ is
closer to unity and hence the present equation of state for the
tachyon field is like that of a pressure-less fluid.  
In such a case accelerating phase of the universe starts at late times
compared to models where $|\phi_{in} H_{in}|$ is closer to its minimum
allowed value.  
Thus we need to fine tune the value of this parameter in order to
arrange for accelerating phase of the universe to start at low
redshifts, as required by observations.

In future evolution, density parameter for matter decreases with the
start of the accelerating phase.  
The density parameter for tachyons begins to increase in this phase and
quickly approaches unity.

Next we study the behavior of these solutions in past, i.e., we use
the present day conditions and evolve back to see the kind of initial
conditions that are required for us to get a viable model today. 
Figures used to illustrate future evolution also show the past
evolution of these quantities.  
As motivated in section \ref{sec:tachmatter}, we expect matter and radiation to drive
$\dot\phi$ away from unity towards smaller values in forward evolution. So,  
as we evolve the equations back, we expect all present
day conditions to lead towards $\dot\phi=1$.  
This is indeed what we find in numerical solutions. The equation of
state for tachyon field approaches that of a pressure-less fluid, and
the universe expands with $a(t) \propto t^{2/3}$ at high redshifts.   
At even earlier times, radiation takes over and both matter and
tachyons become subdominant components.

\begin{figure}
\begin{center}
\epsfig{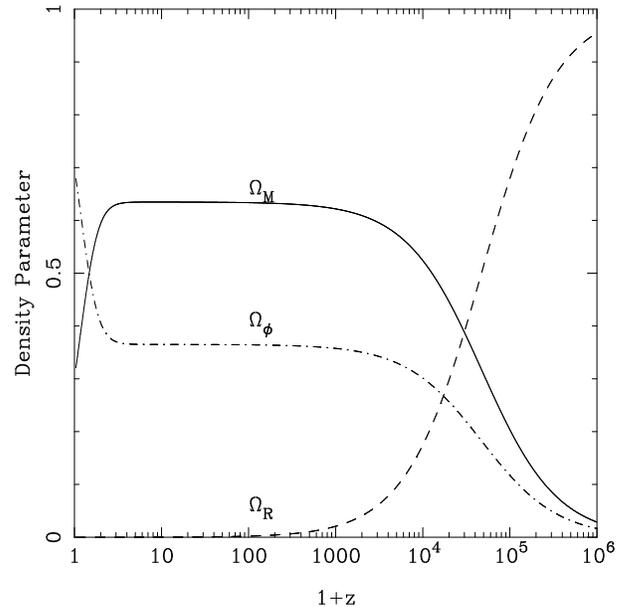}
\end{center}
\caption{$\Omega_M$, $\Omega_{\phi}$ and $\Omega_R$ as  functions of
redshift. The matter density parameter is almost a constant at
$2 \leq z \leq 10^3$ and then drops to small values as radiation begins to
dominate. The present day  values of density parameter for matter and radiation
are $0.3$ and $10^{-5}$ respectively.}
\label{fig:denparm}
\end{figure}

The time dependent matter density parameter $\Omega_M(t) = 8 \pi G \rho_M(t)/ 3
H^2(t) $ is plotted as a function of redshift in
Fig. \ref{fig:denparm}.   
Here non-relativistic matter dominates in the sense that $\Omega_M >
\Omega_{\phi}$ but evolution does not drive $\Omega_{\phi}$ towards
zero or  $\Omega_M$ close to unity.
Instead the two density parameters remain comparable during much of
the ``matter dominated'' era (from $z\approx 10^3$ to $z\approx 2$). 
This is a unique feature  not seen in other cosmological models.
In the phase where the equation of state of tachyon field approaches
$w \simeq 0$, the ratio of density parameter for non-relativistic
matter and tachyon field becomes a constant.  
For models of interest, i.e., where accelerating phase is starting at
low redshifts, this ratio is of order unity for most choices of
parameters and the relative
importance of tachyons and matter does not change up to the time when
accelerating phase begins. 
(This has a significant effect on gravitational instability and growth
of perturbations; see section \ref{sec:tachstr}.)

\begin{figure}
\begin{center}
\epsfig{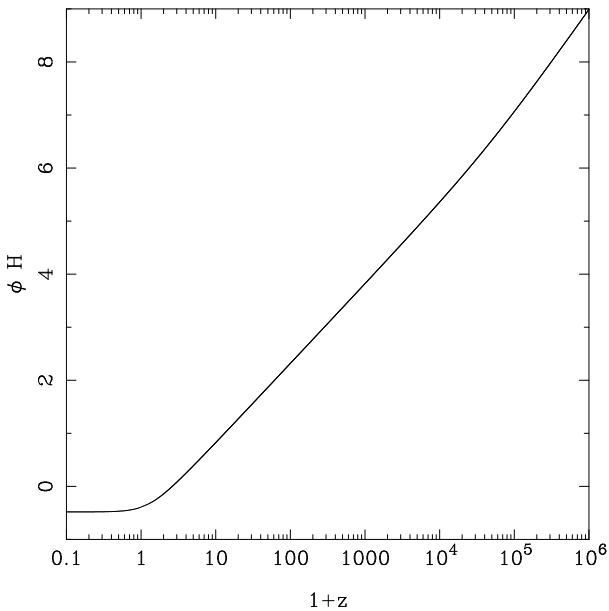}
\end{center}
\caption{Plot of $\phi H$ as a function of redshift.}
\label{fig:phiH}
\end{figure}

The tachyon field $\phi$ approaches a nearly constant value in past
and changes very little from its value in the early 
universe to the time when the accelerating phase begins.  
In the pre-acceleration phase, $\dot\phi$ is close to unity.  This
means that the field $\phi$ starts from large values in  the models that
satisfy basic observational constraints.  
This is the fine tuning required in constructing viable tachyon
models.
In Fig.~\ref{fig:phiH} we show the evolution of $\phi (t) H(t)$ as a
function of redshift.  
As expected, this is a constant in the asymptotic regime where
tachyons dominate.  
However this function evolves rapidly in the dust like regime.
As $\phi$ remains a constant in this regime, the change is mainly due
to change in $H(t)$.
The asymptotic value of this product is $\phi H(t) \longrightarrow
\sqrt{2n/3}$, so we expect $\phi H(t)$ to decrease from its initial
value and make a transition  to the constant asymptotic value.
We need to choose the initial value of this product so that the
transition from dust like to accelerating phase happens around now. 
As the asymptotic value is $\phi H(t) \longrightarrow \sqrt{2n/3}$,
and this is of order unity for small $n$, the value of $\phi$ in the
early universe has to be large.  
The product $\phi(t_{Pl}) H(t_{Pl})$ is approximately $\phi(t_{0})
H(t_{Pl})$ as $\phi$ does not change by much during its evolution.
Since the  present day $\phi(t_0) H_0$ is of order unity, $\phi(t_{Pl}) H_{Pl}\approx
H_{Pl}/H_{0}$.  
(Here $t_0$ is the present time and $H_0$ is the present value of the
Hubble constant.)
Thus the initial value of $\phi$ sets the time scale  when transition
to asymptotic behavior takes place.   
This fine tuning is similar to that needed in models with the
cosmological constant as the value of $\Lambda$ has to be tuned to
arrange for accelerating phase to start at low redshifts. 

\begin{figure}
\begin{center}
\epsfig{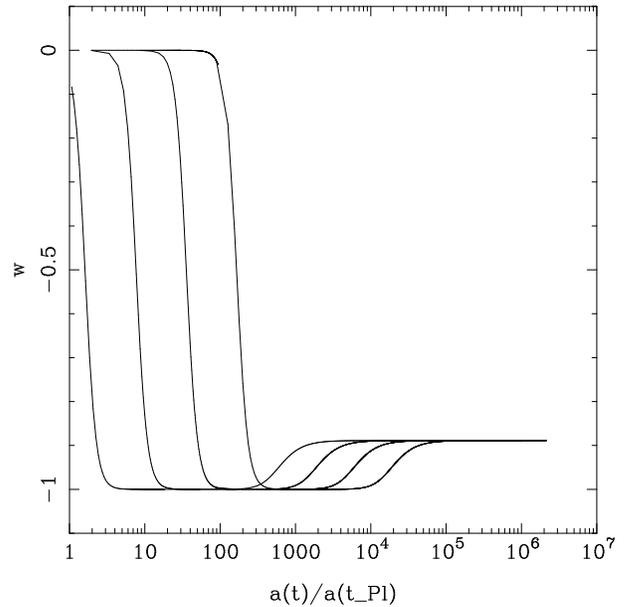}
\end{center}
\caption{Plot of $w_\phi = \dot{\phi}^2 -1$ as a function of the scale
factor. The curve on the left has the lowest value of initial $\phi H$
and increases by an order of magnitude for respective curves as we go
towards right.}
\label{fig:w}
\end{figure}

\begin{figure}
\begin{center}
\epsfig{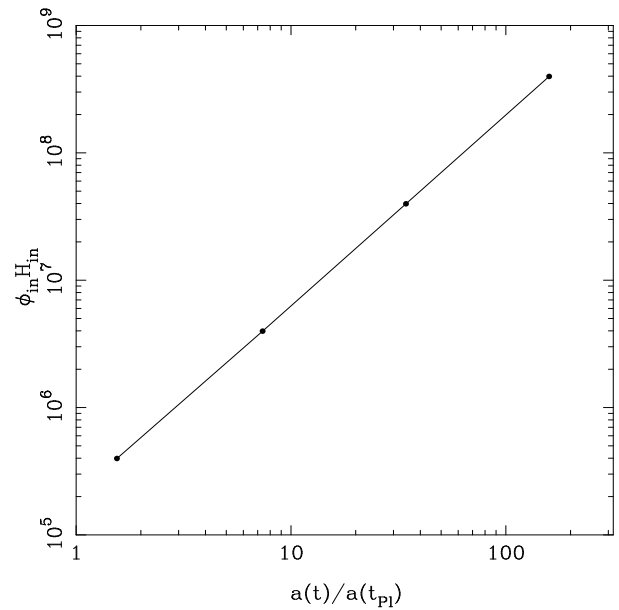}
\end{center}
\caption{Plot of $\phi_{in} H_{in}$ (at Planck time) as a function of
  the scale factor value at which 
$w_\phi =\dot{\phi}^2 -1$ becomes half its asymptotic value.}
\label{fig:wby2}
\end{figure}

To illustrate this point further, we start the integration at Planck
time for a range of initial conditions and show that indeed we obtain
this behavior.  The age of the  universe at the time of transition
from pressure-less 
behaviour for the tachyon field to an effective cosmological
constant  behavior is indeed proportional to the initial value of
$\phi$. 
To illustrate this, we first plot $w$ for the
tachyon field as a function of scale factor in Fig.~\ref{fig:w}. 
This is plotted for many values of $\phi(t_{planck}) H(t_{planck})$. 
An order of magnitude increase in the initial value delays the onset
of accelerating phase by roughly an order. 
Hence to achieve accelerating phase at late times, one requires the
fine tuned value mentioned above.
The same behavior is summarized in
Fig.~\ref{fig:wby2}, where
we plot the scale factor when $w_{\phi} = w_{\phi_{asymptotic}}/2 =
(2/(3n) - 1)/2$ as a function of  $\phi(t_{planck}) H(t_{planck})$.  

\subsection{Exponential Potential}
\label{exponential}

We now  repeat the above analysis for the exponential potential. 
As before, we consider a mix of non-relativistic matter,
radiation/relativistic matter and the tachyon field. 
We fix $\Omega_\phi=0.7$ at the present epoch.  
 (In the literature, it has been suggested that the exponential potential does arise
   in some of the string theoretic models. However, in our approach, we think of 
   $V(\phi)$ as an arbitrary function just as in quintessence models.)
   For the purpose of numerical work, we  have taken the value of $\phi/\phi_0$
   to be $10^2$ at the present epoch.
The choice of $\dot\phi$ fixes the amplitude of the potential as we
have already fixed  $\Omega_\phi$ and $\phi/\phi_0$.
The remaining parameter is $\phi_0$, and as we have already fixed
$\phi_{in}/\phi_0$, fixing $\phi_0$ is equivalent to fixing
$\phi_{in}$.  We use the combination $\phi_{in} H_{in}$ to construct a
dimensionless parameter which indicates the value of $\phi_0$.
$\phi_{in} H_{in}$ should be of order unity if $\phi_{in}$ is
comparable to the age of the Universe today.
We shall study variation of quantities of interest with time for a
range of values of $\dot\phi$ and $\phi_{in} H_{in}$.

In the case of exponential potential, the equation of state for the
tachyon field approaches that of dust as $t\to \infty$.  
Thus any accelerating phase that we might get will be followed by a
dust like phase.  
The  accelerating phase occurs when the tachyon field dominates the
energy density of the universe as it is only in this case that we can
get $\rho + 3p < 0$.  
For an accelerating phase to exist at all,
the tachyon field should begin to dominate {\it before} it enters the
dust like phase.  
Thus the duration of the accelerating phase will depend on how
different $\dot\phi$ is from unity at the start of tachyon dominated
phase. 

\begin{figure}
\begin{center}
\epsfig{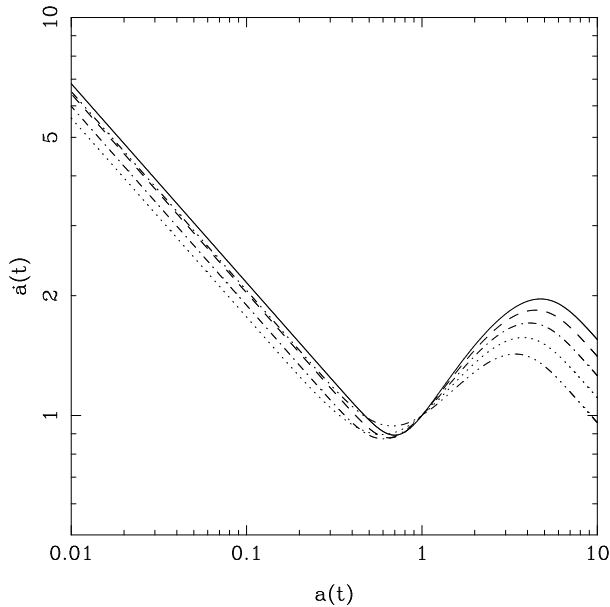}
\end{center}
\caption{Phase plot for the exponential potential. The duration of the
  accelerated phase depends on the 
initial value of $\dot{\phi}$. The accelerated phase ends into a dust
like phase of tachyons.  Solid line is for $\dot{\phi}_{in}=0.1$,
dashed line is for $\dot{\phi}_{in}=0.3$, dot-dashed, dashed and
dot-dot-dashed curves are for $\dot{\phi}_{in}=0.5$, $0.7$ and $0.9$
respectively.} 
\label{fig:phase_expo}
\end{figure}

\begin{figure}
\begin{center}
\epsfig{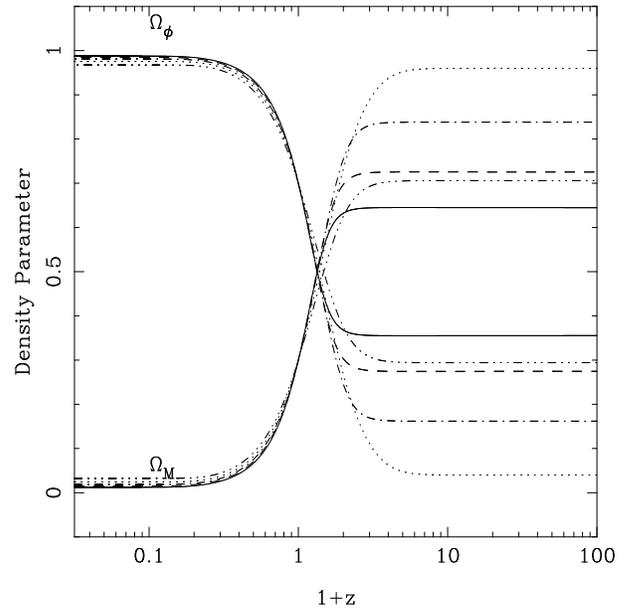}
\end{center}
\caption{Plot of density parameters for matter and tachyon as  functions of
  redshift. The models are the same as in the previous figure.}
\label{fig:omega_expo}
\end{figure}

The phase plot for the scale factor is shown in Fig.~\ref{fig:phase_expo}. 
We have kept $\phi_{in} H_{in}$ fixed at present day and
varied $\dot{\phi}$ for a range of values. 
The unique feature of this model is that we regain the dust like phase
in the future.
It is clear from this plot that the duration of accelerating phase
varies considerably across models shown here.  
The supernova observations require the universe to be in an
accelerating phase at $z < 0.25$, so models that do not have an
accelerating phase at all or too small an accelerating phase can be
ruled out easily.    

The fact that these cosmological models have two decelerating phases,
with an accelerating phase sandwiched in between, is noteworthy. In such
models, one can accommodate the current acceleration of the universe without 
the model ``getting stuck'' in the accelerating phase for eternity. Since the universe
has $a(t)\propto t^{2/3}$ asymptotically, it follows that there
will be no future horizon (for some other attempts to eliminate the
 future horizon, see \cite{nohorz}).   
String theoretic models have difficulty in accommodating an
asymptotically de Sitter-like universe and our model could help in
this context. 

The plot in Fig.~\ref{fig:omega_expo} shows $\Omega_m, \Omega_\phi$ as
functions of redshift and    
shows that $\Omega_\phi$ increases monotonically with time.  

\begin{figure}
\begin{center}
\epsfig{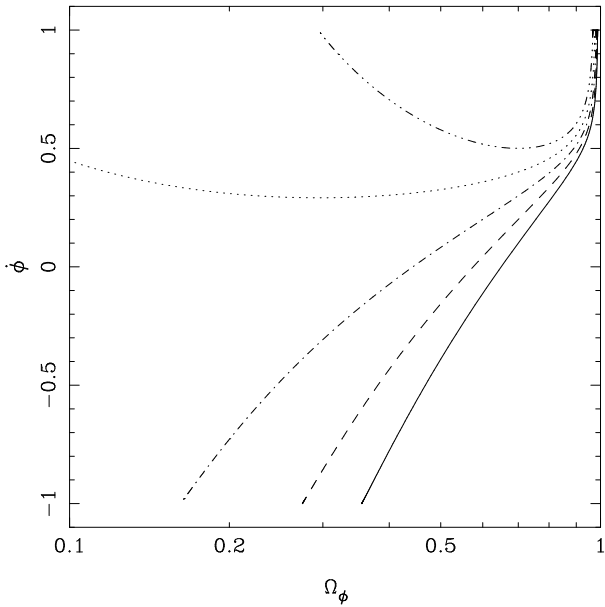}
\end{center}
\caption{Plot of $\dot{\phi}$ Vs. $\Omega_{\phi}$. The asymptotic
  value $\dot{\phi}=1$ is reached irrespective of the initial
  conditions.  The models are the same as in the previous figure.}  
\label{fig:phidot_expo}
\end{figure}

A plot of $\dot{\phi}$ as a function of $\Omega_{\phi}$ in
Fig.~\ref{fig:phidot_expo} shows that all initial conditions lead to
$\dot\phi=1$. 
The function  $\phi H$ increases with  $1+z$.
It is clear that all initial conditions
lead to a narrow range of asymptotic values for this product.

\begin{figure}
\begin{center}
\epsfig{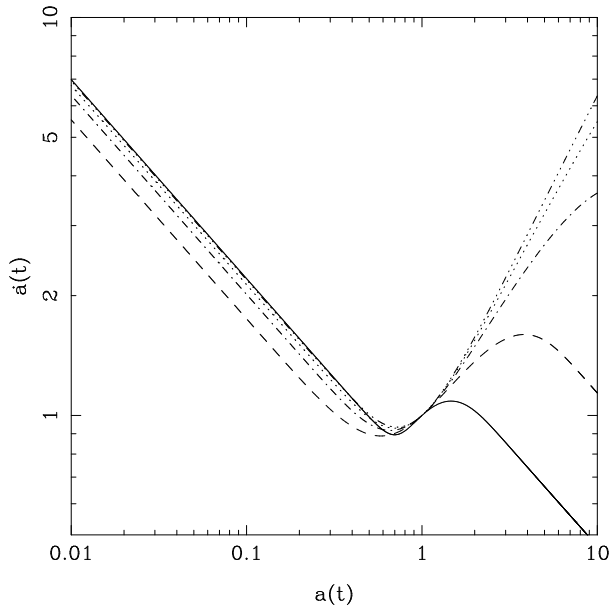}
\end{center}
\caption{Phase plot  for a fixed value
  of $\dot{\phi}=0.38$ and varying $\phi_{in} H_{in}$ (present
  day). The solid line is for $\phi_{in} H_{in}=1/3$, dashed line,
  dot-dashed line, dotted and dot-dot-dashed line for  $\phi_{in}
  H_{in}=   5/3, 7/3, 3$ respectively.   As we increase the  value of
  $\phi_{in} H_{in}$ the   duration    of accelerated phase
  increases.}    
\label{fig:diff_phase}
\end{figure}

\begin{figure}
\begin{center}
\epsfig{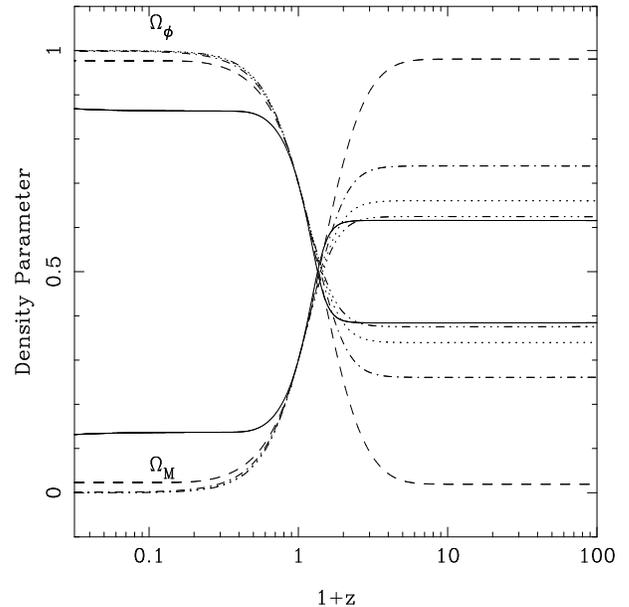}
\end{center}
\caption{This plot shows density parameters for matter and tachyon for a fixed
  value of $\dot{\phi}=0.38$ and varying $\phi_{in} H_{in}$. The
  models are same as in Fig. \ref{fig:diff_phase}.}  
\label{fig:diff_omega}
\end{figure}

If we keep the value of $\dot{\phi}$ fixed and increase $\phi_{in} H_{in}$, the
duration of the accelerating phase increases. 
This is illustrated in Fig.~\ref{fig:diff_phase}.
A plot of the density parameters is shown in Fig.~\ref{fig:diff_omega}.

\section{Comparison with Supernova Ia  observations}
\label{sec:supnova_data}
We now compare the results with supernova Ia observations by
Perlmutter et al. \cite{perl_nova1,perl_nova2}. 
The data from the redshift and luminosity distance observations of
these supernovae, collected by the Supernova Cosmology Project,
concludes that the universe is currently going through an accelerating
phase of expansion.
This is a powerful constraint on cosmological models, e.g. it rules
out the Einstein-deSitter model at a high confidence level.
The obvious contender is a positive  cosmological constant, though
more exotic matter driving the accelerated expansion is also a
possibility.  
Since currently the energy density of the universe is dominated by a
form of matter with negative pressure, the presence of a tachyon source
is thus expected to be consistent with these observations.

The results of theoretical model and observations are compared for
luminosity measured in logarithmic units, i.e., magnitudes defined by 
\begin{equation}
m_{B}(z)={\mathcal M} + 5 log_{10}(D_{L})
\end{equation}
where$ {\mathcal M}=M-5log_{10}(H_0)$ and $D_{L}=H_{0}d_{L}$, the
factor $M$ being the absolute magnitude of the object and $d_L$ is the
luminosity distance 
\begin{equation}
d_{L}=(1+z) a(t_0)r(z);~~~~r(z)=c \int \frac{dt}{a(t)}  
\end{equation}
where $z$ is the redshift.

We perform the $\chi^2$ test of goodness-of-fit on the model with 
$V(\phi) \propto \phi^{-2}$.
For our analysis, we have  used all the 60 supernovae quoted in
\cite{perl_nova1}.
Of these 42 are high red shift supernovae reported by the supernova
cosmology project \cite{perl_nova1,perl_nova2} and 18 low red shift
supernovae of Callan-Tollolo survey \cite{hamuy1,hamuy2}. 
We have three parameters: $\Omega_M(in)$, $n$ and $\phi_{in}
H_{in}$ (with `in' referring to present epoch).  
An additional freedom is the choice of sign of $\dot\phi(x=t_{in}
H_{in})$.
We freeze $\Omega_M(in)$ at $0.3$ and we find that
that for any choice of $n$, we can get a reasonably low
value for reduced $\chi^2$ by choosing $\phi_0 H_0$ judiciously.
In addition, the value of $\chi^2$ does not vary much over the range
of parameter values that we have studied.
Minimum value of $\chi^2$ per degree of freedom that we encountered is
around $1.93$.  This test does not isolate any particular region in the
parameter space so we shall refrain from quoting any particular values
of parameters as our best fit.
Suffice to say that a large range of each of these parameters is
allowed by the supernova observations.
This discussion holds for both the potentials discussed above.

\begin{figure}
\begin{center}
\epsfig{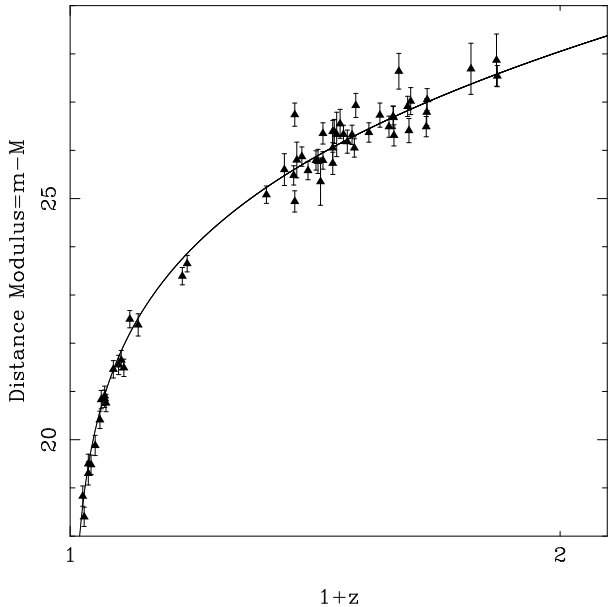}
\end{center}
\caption{Comparison of the model  with $n=6$ and the present $\phi H=2.56$ with
  supernova Ia data.} 
\label{fig:dm}
\end{figure}

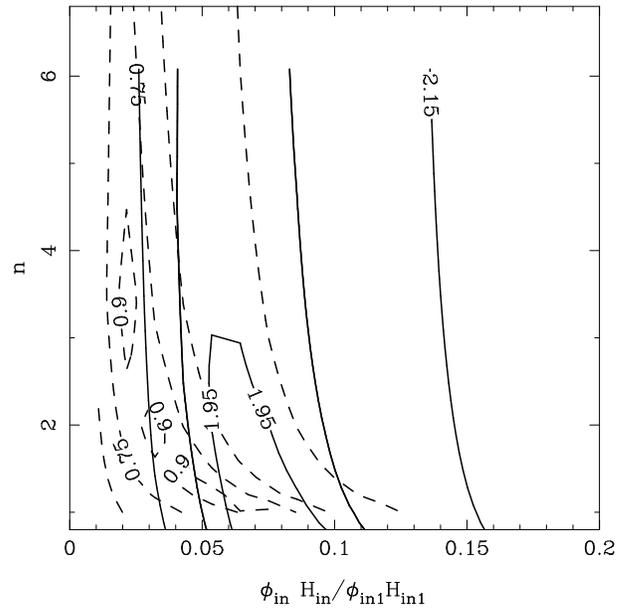
\begin{figure}
\begin{center}
\epsfig{file=chi_phi2.ps,height=8.0cm, width=8.0cm}
\end{center}
\caption{The figure shows contours of $\Omega_M$ at $z=10$ (dashed
  lines) and contours of reduced $\chi^2$ (solid lines) for $V(\phi)
  \propto 1/\phi^2$. There is small region where a  small $\chi^2$
  overlaps $\Omega_m > 0.9$. 
  This region is near $n=1.2$ and $\log(\phi_{in} H_{in}/\phi_{in1}
  H_{in1}) = 0.05$ 
  The values along the x axis are ratios of $\phi_{in} H_{in}$ to the
  minimum value of $\phi_{in} H_{in}$ for a particular value of power
  $n$.}
\label{fig:chi_phi2} 
\end{figure}

We have plotted distance modulus $dm(z) = m - M$ as a function of
redshift for one of the models in Fig.~\ref{fig:dm}. 
The data points for the $60$ supernovae are over plotted as well.
The contours of reduced $\chi^2$ and the value of $\Omega_m$ at redshift $z=10$
are illustrated in Fig.~\ref{fig:chi_phi2}.
This is  of interest while studying
structure formation in these models.  

We also did  a $\chi^2$ analysis of models with a range of parameters for the
exponential potential. 
Here again the theoretical models satisfy the supernova constraints and we
cannot rule out a specific model by this analysis. 
We plot the contours of $\Omega_M$ at $z=10$ and $\chi^2$ in
Fig. \ref{fig:chi_expo}.  
We obtain results similar to those for $1/\phi^2$ potential.

\begin{figure}
\begin{center}
\epsfig{file=chi_expo.ps,height=8.0cm, width=8.0cm}
\end{center}
\caption{Contours of $\Omega_M$ at $z=10$ (dashed lines) and
  contours of reduced $\chi^2$ (solid lines).  The favored region is
  around $\phi_{in} H_{in}=1$ and $0.25 \leq \dot\phi \leq 0.4$.} 
\label{fig:chi_expo}
\end{figure}

\section{Structure formation  in tachyonic models} 
\label{sec:tachstr}

Cosmological models with tachyons and non-relativistic matter have a
significantly different behavior as compared to quintessence or the
cosmological constant models.  
The most important difference here is that the source of acceleration
in the Universe makes an insignificant contribution to the energy
density of the universe beyond $z > 1$ in quintessence or cosmological
constant models, whereas in tachyon models the density parameter for
the scalar field does not become insignificant in comparison with the
density parameter for matter in most models.
This has important implications for structure formation in these
models as the density parameter of the matter that clusters is always
smaller than unity, and the rate at which perturbations grow will be
smaller than in standard models.
The exception to this rule is a small subset of models where
$\Omega_\phi$ approaches values much smaller than unity beyond $z
\approx 1$. 
As can be seen in the contour plots in the previous section, these is
a small subset of the models that satisfy the constraints set by
supernova observations.

Given that the density parameter for matter is almost a constant at 
high redshifts ($3 < z < 10^3$), we can solve for the rate of growth
for density contrast in the linear limit.  
The equation for the density contrast is given by \cite{paddy_prob}
\begin{equation}
\ddot{\delta}+2 \frac{\dot{a}}{a} \dot{\delta}=4 \pi G \rho \delta
\end{equation}
where $\delta=(\rho - \bar{\rho})/ \bar{\rho}$, the factor
$\bar{\rho}$ being the average density.
Rescaling in the same manner as the cosmological equations we have
\begin{equation}
\delta''+2 \frac{H}{H_0} \delta'=\frac{3}{2} \Omega_{M_0}
\frac{a_0^3}{a^3} \delta \label{eq:lindelta}
\end{equation}
Here $\Omega_{M_0}$ is the density parameter for non-relativistic
matter at the epoch when the scale factor is $a_0$ and the Hubble
parameter is $H_0$.
Since, $\Omega_M$ is nearly constant at high redshift, we get $\delta
\propto t^m$, and $m=(1/6)(\sqrt{1 + 24   \Omega_{M}} - 1)$ for
the growing mode. 
The  unique feature here is that at high redshift, matter density
parameter does not saturate at unity for all the models. 
This is true for both exponential and $1/\phi^2$ potentials.
For models where the matter density parameter does not reach unity,
the growth of perturbations is slow, as can be seen from the above
equation. 
The models in which density parameter is unity at high redshifts, the
growth of perturbations is closer to that in the $\Lambda$CDM model.
The slower growth of perturbation implies that {\it rms} fluctuations
in mass distribution were  larger at the time of recombination as
compared to conventional  models.  
This will have an impact on the temperature anisotropies in the
microwave background in these models.
Since the latter is tightly constrained from CMBR measurements, models
with slow growth of perturbations can be ruled out.

\begin{figure}
\begin{center}
\epsfig{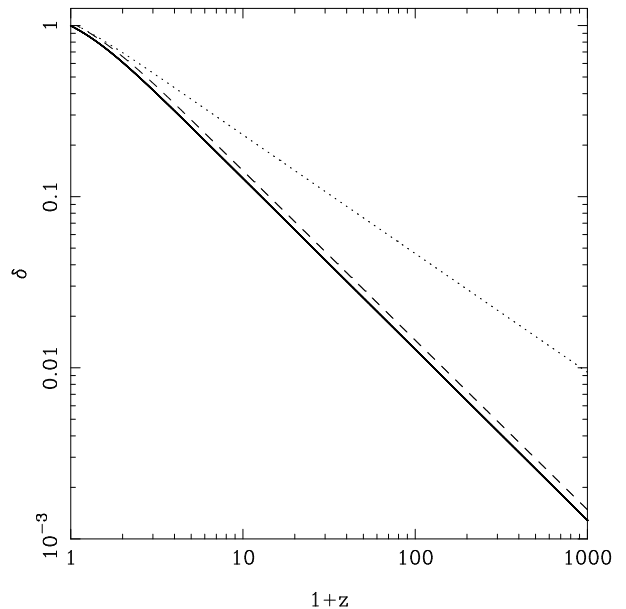}
\end{center}
\caption{Evolution of the density contrast with
  redshift. The curves correspond to $\Omega_{in}=0.3$ (present day
  value). The solid curve is for   $\Lambda$CDM model given for comparison. The  density
  parameter   $\Omega_M \approx 
  0.98$ at $z=10$ for the  dashed line and 
   $\Omega_M \approx 0.56$ at $z=10$ for the dotted line.}  
\label{fig:lndelta}
\end{figure}

We use the solution outlined above to set the initial conditions for
equation (\ref{eq:lindelta}) and evolve forward through the regime where
tachyons begin to dominate and matter becomes irrelevant.  
As the universe begins to accelerate, at late times we anticipate that
the growing mode would slow down and eventually saturate.  
This is indeed true.  
The rate of growth for $\delta$ slows down once the universe begins to
accelerate. 
It comes to a halt around the epoch where the accelerating phase begins.
This late time behavior is similar to what happens in most models
where the universe begins to accelerate at late times. 
The evolution is illustrated in Fig. \ref{fig:lndelta}.
Here we have plotted  two different models, one  in which  density parameter
saturates at a small value and the other in which it approaches unity.

It is clear from the figure that one can indeed construct models in
which the growth of  perturbation is very similar to that in
$\Lambda$CDM models. Such models are clearly viable. It is also
obvious that these models are confined to a narrow range of
parameters, as described in the figure captions; if one moves out of
this range, then the perturbations grow more slowly and should have
higher amplitude in the past in order to maintain a given amplitude
today. These are ruled out by CMBR observations. The following caveat,
however, needs to be kept in mind in ruling out such models. It was
suggested in \cite{tptirth}  that one can construct models with
tachyonic scalar field in which the equation of state is different at
small scales and large scales. In such models, our conclusions will
apply only at large scales and growth of structure at small scales
will still be possible, i.e., inhomogeneities in the tachyon filed
will play and role and may offset the conclusions about growth of
perturbations at small scales whereas our results for the expansion of
the universe will remain valid at sufficiently large scales.

\section{Conclusions}
\label{sec:conclusions}

We have shown that it is possible to construct viable models with
tachyons contributing significantly to the energy density of the
universe.  In these models, matter, radiation and tachyons co-exist.
We show that a subset of these models satisfy the constraints on the
accelerating expansion of the universe.  
For the accelerating phase to occur at the present epoch, it is
necessary to fine tune the initial conditions. 

We have further demonstrated that the density parameter for tachyons
does not becomes negligible at high redshifts, hence the growth of
perturbations in non-relativistic matter is slower for most models than,
e.g., the  $\Lambda$CDM model.  
This problem does not effect a small subset of models.
However, given that the density parameter of tachyons cannot be
ignored in the ``matter dominated era'', it is essential to study the
fate of fluctuations in the tachyon field.

\section*{Acknowledgement}

HKJ thanks Ranjeev Misra for useful discussions.  JSB is grateful to
Rajaram Nityananda for insightful comments.  All the authors thank
K.Subramanian for useful comments.

\end{document}